\documentclass{article}
\usepackage{amssymb,amsmath,amscd, amsthm,stmaryrd,verbatim, mathrsfs}
\usepackage{color}

\newtheorem{Th}{Theorem}
\newtheorem{Def}{Definition}
\newtheorem{Lem}{Lemma}
\newtheorem{claim}{Claim}

\newcommand{\mb}{\mathbf}
\newcommand{\bb}{\mathbb}
\newcommand{\ms}{\mathscr}
\newcommand{\mr}{\mathrm}

\usepackage{hyperref}
\begin{document}

\title{MICZ-Kepler Problems in All Dimensions}
\author{Guowu Meng\\
\small{\it Department of Mathematics, Hong Kong Univ. of Sci. and Tech.}\\
\small{\it Clear Water Bay, Kowloon, Hong Kong}\\
\small{Email: mameng@ust.hk} } \maketitle
\begin{abstract}
The Kepler problem is a physical problem about two bodies which
attract each other by a force proportional to the inverse square of
the distance. The MICZ-Kepler problems are its natural cousins and
have been previously generalized from dimension three to dimension
five. In this paper, we construct and analyze the (quantum)
MICZ-Kepler problems in all dimensions higher than two.
\end{abstract}

%\tableofcontents

\section {Introduction}
The Kepler problem is the physics problem about two bodies which
attract each other by a force proportional to the inverse square of
the distance. By solving this problem in classical mechanics, Newton
gave a satisfactory explanation for Kepler's laws for the planetary
motion. The Kepler problem plays a significant role in the
development of quantum mechanics, too; in fact, the solution of this
problem in the Schr\"{o}dinger's wave mechanics firmly puts the
Schr\"{o}dinger equation right at the center of quantum mechanics.

After more than three centuries, the Kepler problem still plays an
important role in mathematics and physics. There has been a
continuous interest in this problem; in particular, in the last
three decades we have witnessed an explosion of its interactions
with quantum mechanics, celestial mechanics and mathematics.  For a
recent comprehensive treatment of the Kepler problem, the interested
readers may consult Ref. \cite{Cordani2002}.

The MICZ-Kepler problems are natural cousins of the Kepler problem,
and they were independently discovered by McIntosh-Cisneros
\cite{MC70} and Zwanziger \cite{Z68} more than thirty years ago.
Roughly speaking, a MICZ-Kepler problem is the Kepler problem in the
case when the nucleus of a hypothetic hydrogen atom also carries a
magnetic charge. These generalized problems share the following
characteristic beauty with the Kepler problem: the existence of the
Runge-Lenz vector and the dynamical $\hbox{Spin}(4)$ symmetry for
the bound states; therefore, they provide a rich family of examples
for the exploration of extra hidden dynamic symmetry.

The hamiltonian of a MICZ-Kepler problem is constructed from that of
the Kepler problem by adding the vector potential of a Dirac
monopole and a repulsive centrifugal potential; explicitly, we have
\begin{eqnarray}
H = \frac{1}{2m}(\vec p+e\vec A)^2 + \frac{\mu^2}{2m r^2}
 - \frac{e^2}{r}
\end{eqnarray}
where $\vec p$ is the canonical momentum of the electron, $\vec A$
is the vector potential of a Dirac monopole, $r$ is the distance
from the electron to the hydrogen nucleus, $m$ is the (reduced) mass
of the electron, $e$ is the fundamental unit of the electric charge,
and $\mu$ is the magnetic charge of the Dirac monopole measured in
unit $c\over e$, i.e.,  $\mu{c\over e}$ is  the magnetic charge of
the Dirac monopole, here $c$ is the speed of light in
vacuum\footnote{The Dirac quantization condition becomes ${\mu\over
\hbar}=$ a half integer.}.

Quantum mechanically, via rescaling
\begin{eqnarray}
r\to {\hbar^2\over m e^2}r, \quad \mu\to \hbar \mu,\nonumber
\end{eqnarray}we arrive at the following hamiltonian operator:
\begin{eqnarray}
{\hat H}
 = {me^4\over \hbar^2}\left(-\frac{1}{2}{\Delta}_A + \frac{\mu^2}{2r^2}
 - \frac{1}{r}\right):= {me^4\over \hbar^2}\hat h\nonumber
\end{eqnarray}
where\footnote{Remark that $\hat h$ is dimensionless and is
expressed in terms of dimensionless quantities. It is $\hat h$ (not
$\hat H$) that will be generalized later. }
\begin{eqnarray}\label{KH}
{\hat h}
 = -\frac{1}{2}{\Delta}_A + \frac{\mu^2}{2r^2}
 - \frac{1}{r}\;.
\end{eqnarray}
Here $\Delta_A$ is the Laplace operator twisted by the gauge
potential $A$ of a Dirac monopole, and $\mu=0$, $\pm {1\over 2}$,
$\pm 1$, $\cdots$ is the magnetic charge of the Dirac monopole
measured in terms of the fundamental unit\footnote{The case $\mu=0$
corresponds to the Kepler problem.}. Locally, with a gauge chosen,
we have
\begin{eqnarray*} \Delta_A=\nabla_a\nabla_a
\end{eqnarray*}
where the repeated index $a$ is summed up. Here $\nabla_a$ is the
$a$-th covariant partial derivative and is written as
$\partial_a+iA_a$ by physicists with $\vec A=(A_1, A_2, A_3)$ being
the gauge potential of the Dirac magnetic monopole. Mathematically
$\nabla_a=\partial_a+\omega_a$ where $\omega=\omega_a dx_a$ has been
previously identified with the Levi-Civita spin connection form of
the cylindrical metric $$ds^2={1\over r^2}(dx_1^2+dx_2^2+dx_3^2)$$
on the punctured $3$-space, see Ref. \cite{meng04} for the details.

The MICZ-Kepler problems exist in higher dimensions just as the
Kepler problem does, and that is a main observation here. In fact,
the existence in dimension five has been previously observed
\cite{Iwai90}; however, the existence in all dimensions greater than
two, though very straightforward from a canonical geometric point of
view, was probably not expected by the community. This overlook is
very likely due to a general belief in the literature: the existence
of Dirac monopoles and its five dimensional analogue (the Yang
monopoles \cite{yang78}) has to do with the existence of the
division algebras or Hopf bundles.

\vskip 10pt In section 2, we construct the MICZ-Kepler problems in
all dimensions and then state the main results. The construction is
geometric and canonical. A key ingredient in the construction is the
higher dimensional generalization of the Dirac monopoles
--- a canonical geometric object that has been used in Ref.
\cite{meng03}. In section 3, we first introduce the explicit
formulas for the gauge potential of these generalized Dirac
monopoles; then we list and prove some crucial identities necessary
for the exhibition of the extra large hidden dynamical symmetry. In
section 4, we introduce the angular momentum and Rung-Lenz vector
for our MICZ-Kepler problems, and derive the symmetry algebra. In
section 5, we obtain the energy spectrum and the energy eigenspaces
for bound states by using Painlev\'{e} analysis plus representation
theory, and then show that the Hilbert space of bound states has a
hidden dynamical $\hbox{Spin}(D+1)$-symmetry for a $D$-dimensional
MICZ-Kepler problem even though in general the Runge-Lenz vector
fails to be conserved when $D$ is even.

\vskip 10pt Remark that the MICZ-Kepler problems in higher
dimensions constructed here are based on modern geometry, but they
are solved by classical analytic method with the help of the
representation theory for Lie groups. The solution of these new
MICZ-Kepler problems can in principle be solved by the modern
geometric quantization approach pioneered by Simms \cite{Simms73},
Mladenov and Tsanov \cite {MT85, MT87}, but that will be reserved
for the future for the following reasons: 1) the primary objective
of this paper is to inform the experts in the fields that the
MICZ-Kepler problems do exist in higher dimensions, 2) the classical
analytic approach is more elementary and easier to understand, 3)
the modern geometric quantization approach is a bit more involved
and deserves an independent research.

\subsection*{Acknowledgment}
I would like to thank SiXia Yu for a conversation on the Kepler
problem. This work is supported by the Hong Kong Research Grants
Council under the RGC project no. 602504. I would also like to thank
the referee for his or her careful reading of the manuscript and for
his or her valuable suggestions.

\section{The main results}
From the physics point of view, a MICZ Kepler problem is obtained
from the Kepler problem by adding a suitable background magnetic
field, while at the same time making a suitable adjustment to the
scalar Coulomb potential so that the problem is still integrable.
The background magnetic field is just the spin
connection\footnote{For readers without sufficient background in
modern geometry, just take our explicit formulas for gauge potential
in Eq. (\ref{mnple}) for granted. } of the cylindrical metric on the
configuration space that we have mentioned in the introduction. The
configuration space is the punctured Euclidean space. With this in
mind, we are now ready to give the detailed presentation of our
generalized MICZ Kepler problems.

\vskip 10pt Let $D\ge 3$ be an integer, $\bb R^{D}_*$ be the
punctured $D$-space, i.e., $\bb R^D$ with the origin removed. Let
$ds^2$ be the cylindrical metric on $\bb R^D_*$. Then $(\bb R_*^D,
ds^2)$ is the product of the straight line $\bb R$ with the round
sphere ${\mr S}^{D-1}$. When $D$ is odd, we let $\cal S_\pm$ be the
positive/negative spinor bundle of $(\bb R^D_*, ds^2)$, and when $D$
is even, we let $\cal S$ be the spinor bundle of $(\bb R^D_*,
ds^2)$. Note that, these bundles correspond to the fundamental spin
representations ${\bf s}_\pm$ of $\mr{so}(even)$ and ${\bf s}$ of
$\mr{so}(odd)$ respectively.

The above spinor bundles come with a natural $\mr{SO}(D)$ invariant
connection
--- the Levi-Civita spin connection of $(\bb R^D_*, ds^2)$. As a
result, the Young product of $I$ copies of these bundles, denoted by
${\cal S}_+^I$, ${\cal S}_-^I$ (when $D$ is odd) and ${\cal S}^I$
(when $D$ is even) respectively, come with a natural connection,
too.

For the sake of notational sanity, from here on, when $D$ is odd and
$\mu$ is a half integer, we rewrite ${\cal S}_+^{2\mu}$ as ${\cal
S}^{2\mu}$ if $\mu\ge 0$ and rewrite ${\cal S}_-^{-2\mu}$ as ${\cal
S}^{2\mu}$ if $\mu\le 0$; moreover, we adopt this convention for
$\mu=0$: ${\cal S}^{0}$ is the product complex line bundle with the
product connection. When $D$ is odd,  ${\cal S}^{2\mu}$ is the
product complex line bundle with the product connection in the case
$\mu=0$, and is the fundamental spinor bundle $\cal S$ in the case
$\mu=1/2$.

Note that ${\cal S}^{2\mu}$ is our analogue of the Dirac monopole
with magnetic charge $\mu$, and its corresponding representation of
$\mr{so}(D-1)$ will be denoted by ${\bf s}^{2\mu}$. We are now ready
to present our definitions.

\begin{Def} Let $n\ge 1$ be an integer, $\mu$ a half integer.
The $(2n+1)$-dimensional MICZ-Kepler problem with magnetic charge
$\mu$ is defined to be the quantum mechanical system on $\bb
R^{2n+1}_*$ for which the wave-functions are sections of ${\cal
S}^{2\mu}$, and the hamiltonian is
\begin{eqnarray}
{\hat h}
 = -\frac{1}{2}\Delta_\mu + \frac{(n-1)|\mu|+\mu^2}{2r^2}
 - \frac{1}{r}
\end{eqnarray}
where $\Delta_\mu$ is the Laplace operator twisted by ${\cal
S}^{2\mu}$.
\end{Def}

\begin{Def} Let $n>1$ be an integer, $\mu=0$ or $1/2$.
The $2n$-dimensional MICZ-Kepler problem with magnetic charge $\mu$
is defined to be the quantum mechanical system on $\bb R^{2n}_*$ for
which the wave-functions are sections of ${\cal S}^{2\mu}$, and the
hamiltonian is
\begin{eqnarray}
{\hat h}
 = -\frac{1}{2}\Delta_\mu + \frac{(n-1)\mu}{2r^2}
 - \frac{1}{r}
\end{eqnarray}
where $\Delta_\mu$ is the Laplace operator twisted by ${\cal
S}^{2\mu}$.
\end{Def}
Note that we require $\mu=0$ or $1/2$ in the even dimensional case.
There is both an analytic and an algebraic reason for this
requirement, which shall be pointed out in appropriate places.
Remark also that, upon a choice of a local gauge, the background
magnetic potential $A_\alpha$ can be explicitly written down, then
$\Delta_\mu=\sum_\alpha(\partial_\alpha +i A_\alpha)^2$ can be
explicitly written down, too. We are now ready to state our main
results.
\begin{Th}  Let $n \ge 1$ be an integer and $\mu$ be a half integer. For the $(2n+1)$-dimensional
MICZ-Kepler problem with magnetic charge $\mu$, the following
statements are true:

1) The negative energy spectrum is
$$
E_I=-{1/2\over (I+n+|\mu|)^2}
$$ where $I=0$, $1$, $2$, \ldots;

2) The Hilbert space $\ms H$ of negative-energy states admits a
linear $\mr{Spin}(2n+~2)$-action under which there is a
decomposition
$$
{\ms H}=\hat\bigoplus _{I=0}^\infty\,{\ms H}_I
$$ where ${\ms H}_I$ is a model for the irreducible $\mr{Spin}(2n+2)$-representation
with highest weight $(I+|\mu|,|\mu|, \cdots, |\mu|, \mu)$;

3) $\mr{Spin}(2n+1,1)$ acts linearly on the positive-energy states
and $\mr{Spin}(2n+~1)\rtimes~ {\bb R}^{2n+1}$ acts linearly on the
zero-energy states;

4) The linear action in either part 2) or part 3) extends the
manifest linear action of $\mr{Spin}(2n+1)$, and ${\ms H}_I$ in part
2) is the energy eigenspace with eigenvalue $E_I$ in part 1).
\end{Th}

\begin{Th}  Let $n>1$ be an integer and $\mu=0$ or $1/2$.  For the $2n$-dimensional
MICZ-Kepler problem with magnetic charge $\mu$, the following
statements are true:

1) The negative energy spectrum is
$$
E_I=-{1/2\over (I+n+\mu-{1\over 2})^2}
$$ where $I=0$, $1$, $2$, \ldots;

2) The Hilbert space $\ms H$ of negative-energy states admits a
linear $\mr{Spin}(2n+~1)$-action under which there is a
decomposition
$$
{\ms H}=\hat\bigoplus _{I=0}^\infty\,{\ms H}_I
$$ where ${\ms H}_I$ is a model for the irreducible $\mr{Spin}(2n+1)$-representation
with highest weight $(I+\mu,\mu, \cdots, \mu)$;

3) $\mr{Spin}(2n,1)$ acts linearly on the positive-energy states and
$\mr{Spin}(2n)\rtimes {\bb R}^{2n}$ acts linearly on the zero-energy
states;

4) The linear action in part 2) extends the manifest linear action
of $\mr{Spin}(2n)$, and ${\ms H}_I$ in part 2) is the energy
eigenspace with eigenvalue $E_I$ in part 1).

\end{Th}
Remark that, based on the analysis done in later sections, we know
that bound eigen-states are always the ones with negative energy
eigenvalues.

\section {Generalized Dirac monopoles}
We write $\vec r = (x_1, x_2, \ldots, x_{D-1}, x_0)$ for a point
in $\bb R^{D}$ and $r$ for the length of $\vec r$. The small Greek
letters $\mu$, $\nu$, etc run from $0$ to $D-1$ and the small
Latin letters $a$, $b$ etc run from $1$ to $D-1$. We use the
Einstein convention: the repeated index is always summed up.

To do computations, we just need to choose a gauge on $\bb R^{D}$
minus the negative $0$-th axis and then write down the gauge
potential explicitly. We have done that before in Eq. (10) of Ref.
\cite{meng04}.  Note that, if we use the rectangular coordinates
$\vec r =(\vec x, x_0)$, then the gauge potential $A=A_\mu dx_\mu$
from Eq. (10) of Ref. \cite{meng04} can be written as\footnote{In
Ref. \cite{meng04} we only consider the case $D$ is odd
--- the topological nontrivial case, but the basic construction
there is valid in any dimension, see appendix A in Ref.
\cite{meng03}. }
\begin{eqnarray}\label{mnple}
A_0=0,\hskip 20 pt A_b=-{1\over r(r+x_0)}x_a\gamma_{ab}
\end{eqnarray}
where $\gamma_{ab}={i\over 4}[\gamma_a,\gamma_b]$ with $\gamma_a$
being the ``gamma matrix" for physicists.  Note that $\gamma_a=ie_a$
with $\vec e_a$ being the element in the Clifford algebra that
corresponds to the $a$-th standard coordinate vector of $\bb
R^{D-1}$.

It is straightforward to calculate the gauge field strength
$F_{\mu\nu}=
\partial_\mu A_\nu-\partial_\nu A_\mu + i [A_\mu, A_\nu]$ and get
\begin{eqnarray}F_{0b}& = & {1\over r^3}x_a\gamma_{ab}\\
F_{ab}& = & -{2\gamma_{ab}\over r(r+x_0)}+  {1\over
r^2(r+x_0)^2}\cdot\cr & &\left((2+{x_0\over r})x_c(x_a
\gamma_{cb}-x_b \gamma_{ca}) +ix_d x_c[\gamma_{d a},\gamma_{c
b}]\right)\end{eqnarray}

The following lemma is crucially used when we check the dynamical
symmetry of our models.
\begin{Lem}\label{lemma}
For the gauge potential defined in Eq. (\ref{mnple}), we have

1) Let $\nabla_\alpha=\partial_\alpha+iA_\alpha$, then the following
identities are valid in any representation:
\begin{eqnarray}\label{Id}
F_{\mu\nu}F^{\mu\nu}=\frac{2}{r^4}c_2\hskip 10pt \hbox{where
$c_2=c_2[\mr{so}(D-1)]={1\over 2}\gamma_{ab}\gamma_{ab}$}
\\
{[\nabla_\kappa,F_{\mu\nu}]}={1\over r^2}\left( x_\mu
F_{\nu\kappa}+x_\nu F_{\kappa \mu}-2x_\kappa F_{\mu\nu} \right)
\\x_\mu A_\mu=0,\hskip 20pt x_\mu F_{\mu\nu}=0, \hskip 20pt [\nabla_\mu,
F_{\mu\nu}]=0
\\r^2[F_{\mu\nu},
F_{\alpha\beta}]+iF_{\mu\beta}\delta_{\alpha\nu}-
iF_{\nu\beta}\delta_{\alpha\mu}+iF_{\alpha\mu}\delta_{\beta\nu}-iF_{\alpha\nu}\delta_{\beta\mu}\cr
={i\over r^2}\left(x_\mu x_\alpha F_{\beta\nu}+x_\mu x_\beta
F_{\nu\alpha}-x_\nu x_\alpha F_{\beta\mu}-x_\nu x_\beta
F_{\mu\alpha} \right)
\end{eqnarray}

2) When $D=2n+1$, identity
\begin{eqnarray}
r^2F_{\lambda\alpha}F_{\lambda\beta}={c_2\over n}\left({1\over
r^2}\delta_{\alpha\beta}-{x_\alpha x_\beta\over
r^4}\right)+i(n-1)F_{\alpha\beta}
\end{eqnarray} holds in the
irreducible representation ${\bf s}^{2\mu}$ of $\mr{so}(2n)$ whose
highest weight is of the form $(|\mu|, \cdots, |\mu|, \mu)$.

3) When $D=2n$, identity
\begin{eqnarray}\label{idevencase}
r^2F_{\lambda\alpha}F_{\lambda\beta}={n-1\over 2}\left({1\over
r^2}\delta_{\alpha\beta}-{x_\alpha x_\beta\over
r^4}\right)+i(n-{3\over 2})F_{\alpha\beta}
\end{eqnarray} holds in the fundamental spin representation $\bf s$ of $\mr{so}(2n-1)$.

\end{Lem}
One can show that the Eq. (\ref{idevencase}) is valid only when
$\mu=0$ or $1/2$. That is the algebraic reason for requiring $\mu=0$
or $1/2$.

\subsection{Proof of Lemma \ref{lemma}} The verification of these
identities is just a direct and lengthy calculation. However, if we
exploit the symmetry, we just need to check the identities at point
$\vec r_0=(0,\ldots, 0, r)$, a much easier task. For example, since
\begin{eqnarray} A_\mu =0, \hskip 20pt F_{0a}=0, \hskip 20pt
F_{ab}=-{1\over r^2}\gamma_{ab}\end{eqnarray} at $\vec r_0$,
identity (\ref{Id}) is obvious.

{\bf {Proof of part 1)}}. We have just remarked that identity
(\ref{Id}) is obvious. Also,
$$
x_\mu F_{\mu\nu}|_{\vec r_0}=x_0F_{0\nu}|_{\vec r_0}=0.
$$
It is also easy to see that $x_\mu A_\mu|_{\vec r_0}=0$ and
$$
[\nabla_\mu, F_{\mu\nu}]|_{\vec r_0}=\partial_\mu F_{\mu\nu}|_{\vec
r_0}=0.
$$
Therefore, identity (10) is checked.

To check identity (9), first we assume $\mu=0$, $\nu=b$, then we
need to check that
$$
\partial_\kappa F_{0b}={1\over r}F_{b\kappa}
$$
at $\vec r_0$, and that can be easily seen to be true whether
$\kappa=0$ or $a$. Next we assume that $\mu=a$ and $\nu=b$, then we
need to check that
$$
\partial_\kappa F_{ab}=-{2\over r^2}x_\kappa F_{ab}
$$
at $\vec r_0$, and that can be easily verified, too.

We divide the checking of identity (11) at $\vec r_0$ into two
cases: 1) one of indices is zero, easy to check;  2) none of the
indices is zero, then the identity becomes
\begin{eqnarray}\label{commutatorSO(EVEN)}
-[\gamma_{ab},\gamma_{cd}]=-i\gamma_{ad}\delta_{bc}+i\gamma_{bd}\delta_{ac}-i\gamma_{ca}\delta_{bd}+i\gamma_{cb}\delta_{ad}
\end{eqnarray} which is of course true because $\gamma_{ab}$'s are the generators of $\mr {so}(D-1)$.

{\bf {Proof of part 2)}}. Write
$$
r^2F_{\lambda\alpha}F_{\lambda\beta}={r^2\over
2}\{F_{\lambda\alpha}, F_{\lambda \beta}\}+{r^2\over
2}[F_{\lambda\alpha}, F_{\lambda \beta}].
$$ Using identity (12), we have
$$
r^2F_{\lambda\alpha}F_{\lambda\beta}={r^2\over
2}\{F_{\lambda\alpha}, F_{\lambda \beta}\}+i(n-1)F_{\alpha\beta}.
$$
Therefore, by checking at $\vec r_0$, we just need to verify that
identity
\begin{eqnarray}\label{specialodd}
\sum_k\{\gamma_{ki}, \gamma_{kj}\}={\delta_{ij}\over n}\sum_{a,
b}(\gamma_{ab})^2
\end{eqnarray}
holds in the irreducible representation ${\bf s}^{2\mu}$ of
$\mr{so}(2n)$ whose highest weight is of the form $(|\mu|, \cdots,
|\mu|, \mu)$. Since this checking is a bit involved, we do it in the
appendix.

{\bf {Proof of part 3)}}.  Write
$$
r^2F_{\lambda\alpha}F_{\lambda\beta}={r^2\over
2}\{F_{\lambda\alpha}, F_{\lambda \beta}\}+{r^2\over
2}[F_{\lambda\alpha}, F_{\lambda \beta}].$$ Using identity (12),
we have
$$
r^2F_{\lambda\alpha}F_{\lambda\beta}={r^2\over
2}\{F_{\lambda\alpha}, F_{\lambda \beta}\}+i(n-{3\over
2})F_{\alpha\beta}.
$$
Therefore, by checking at $\vec r_0$, we just need to verify that
identity
\begin{eqnarray}\label{specialeven}
\sum_k\{\gamma_{ki}, \gamma_{kj}\}=(n-1)\delta_{ij}
\end{eqnarray}
holds in the spin representation $\bf s$ of $\mr{so}(2n-1)$, but
this is easy to check by using the Clifford algebra.

\section {The hidden dynamical symmetry}
To exhibit the dynamical symmetry for our MICZ-Kepler problems, as
usual, we introduce the angular momentum tensor
\begin{eqnarray}
{\hat L}_{\alpha\beta} = -i(x_\alpha{\nabla}_\beta
-x_\beta{\nabla}_\alpha)+r^2F_{\alpha\beta}
\end{eqnarray}
and the Runge--Lenz vector
\begin{eqnarray}\label{Lentz}
{\hat L}_\beta = -\frac{i}{2}\left({\nabla}_\alpha{\hat
L}_{\alpha\beta}+ {\hat
L}_{\alpha\beta}{\nabla}_\alpha\right)+\frac{x_\beta}{r}
\end{eqnarray}
With the help of the identities stated in lemma \ref{lemma}, a
lengthy calculation yields the following commutation relations:
\begin{eqnarray}\label{CTR}
\fbox{$\begin{array}{lcl} [\hat L_{\mu\nu}, \hat h] & = & 0\cr [\hat
L_{\mu\nu}, {\hat L}_{\alpha\beta}] & = & i\delta_{\mu\alpha}{\hat
L}_{\nu\beta}-i\delta_{\nu\alpha}{\hat L}_{\mu\beta}-
i\delta_{\mu\beta}{\hat L}_{\nu\alpha}+i\delta_{\nu\beta}{\hat
L}_{\mu\alpha}\cr [\hat L_{\mu\nu}, \hat L_\lambda] &= &
i\delta_{\mu\lambda}\hat L_\nu-i\delta_{\nu\lambda}\hat L_\mu\cr
[\hat L_\mu, \hat h] & = & 0\cr[\hat L_\mu, \hat L_\nu] & = &
-2i\hat h\hat L_{\mu\nu}.
\end{array}$}
\end{eqnarray}

\vskip 10pt On the Hilbert space of negative-energy states, we can
introduce $\hat J_{MN}$ where the capital Latin letters $M$, $N$ run
from $0$ to $D$:
\begin{eqnarray*}
\hat J_{MN}=\left\{\begin{array}{ll} \hat L_{\mu\nu} & \hbox{if
$M=\mu$, $N=\nu$}\cr (-2\hat h)^{-{1\over 2}}\hat L_\mu & \hbox{if
$M=\mu$, $N=D$}\cr -(-2\hat h)^{-{1\over 2}}\hat L_\nu & \hbox{if
$M= D$, $N=\nu$}\cr 0 & \hbox{if $M=N$}\end{array}\right.
\end{eqnarray*}
Then the commutation relation (\ref{CTR}) says that a
$D$-dimensional MICZ-Kepler problem has a dynamical
$\hbox{SO}(D+1)$-symmetry on the Hilbert space of negative-energy
states. Actually, the dynamical symmetry group should be
$\hbox{Spin}(D+1)$, rather than $\hbox{SO}(D+1)$. It is also clear
that a $D$-dimensional MICZ-Kepler problem has a dynamical
$\hbox{Spin}(D, 1)$-symmetry on the positive-energy states and a
dynamical $\hbox{Spin}(D)\rtimes {\bb R}^D$-symmetry on the
zero-energy states.

It also follows from (\ref{CTR}) that $\hat h$ must be in the center
of Lie algebra $\hbox{so}(D+1)$; in fact, it is a function of the
quadratic Casimir operator of $\hbox{so}(D+1)$:
\begin{eqnarray}\label{hoperator}
\hat h=-{1/2\over c_2[\mr{so}(D+1)]+({D-1\over 2})^2- \bar c_2}
\end{eqnarray} where $\bar c_2$ is the value of $c_2[\mr{so}(D-1)]$
in representation ${\bf s}^{2\mu}$.

To prove Eq. (\ref{hoperator}), we first note that $\hat
L_\mu=(-2\hat h)^{1\over 2}\hat J_{\mu D}$, so $\hat L_\mu \hat
L_\mu=-2\hat h\sum_\mu \hat J_{\mu D}\hat J_{\mu D}$. On the other
hand, based on the definition of $L_\mu$ given in Eq. (\ref{Lentz}),
a direct computation yields
\begin{eqnarray*}
\hat L_\mu \hat L_\mu=1+\left({1\over 2}(D-1)^2-2\bar c_2+\hat
J_{\mu\nu}\hat J_{\mu\nu}\right) \hat h.
\end{eqnarray*}
Therefore,
$$1+\left({1\over 2}(D-1)^2-2\bar c_2+\hat J_{MN}\hat J_{MN}\right)\hat h=0,$$
then we have Eq. (\ref{hoperator}).

\vskip 10pt It is clear now that in order to determine the spectrum
of $\hat h$, we just need to find out which irreducible
representation of $\hbox{Spin}(D+1)$ enters into the Hilbert space
$\ms H$  of negative-energy states. However, we shall find the
discrete spectrum by solving the Schr\"{o}dinger equation directly
and then figure out the decomposition of the Hilbert space of
negative-energy states into the irreducible representations of
$\hbox{Spin}(D+1)$ via representation theory.

\section {The spectrum analysis}

The Schr\"{o}dinger equation for the stationary states, in terms of
the polar coordinates, is
\begin{eqnarray}
\left(-{1\over 2r^{D-1}}\partial_r
r^{D-1}\partial_r+{c_2[\hbox{so}(D)]-\bar c_2+\delta_D\over
2r^2}-{1\over r}\right)\psi=E\psi
\end{eqnarray}
where $E$ is the energy, $c_2[\hbox{so}(D)]={1\over 2}\hat
L_{\mu\nu}\hat L_{\mu\nu}$, and $\delta_D$ is equal to ${(n-1)\mu}$
if $D=2n$ and is equal to $(n-1)|\mu|+\mu^2$ if $D=2n+1$. Under the
action of $\hbox{Spin}(D)$, the Hilbert space of negative-energy
states splits into the direct sum of irreducible components. These
irreducible components are essentially labeled by a nonnegative
integer $l$, and we shall be able to see that shortly. On the
irreducible component labeled by $l$, the Schr\"{o}dinger equation
becomes an equation for the radial part:
\begin{eqnarray}
\left(-{1\over 2r^{D-1}}\partial_r r^{D-1}\partial_r+{c_2[l]-\bar
c_2+\delta_D\over 2r^2}-{1\over r}\right)R_{kl}=E_{kl}R_{kl}
\end{eqnarray} where $c_2[l]$ is the value of
the quadratic Casimir operator of $\hbox{so}(D)$ in the irreducible
component labeled by $l$, and the additional label $k$ is introduced
for the purpose of listing the radial eigenfunctions, just as in the
Kepler problem.

Let $E_{kl}=-{1\over 2}\lambda_{kl}^2$ and
$R_{kl}(r)=e^{-\lambda_{kl}r}u_{kl}$, then the preceding radial
Schr\"{o}dinger equation becomes
\begin{eqnarray}\label{reqn}
\left(-{1\over 2r^{D-1}}\partial_r
r^{D-1}\partial_r+\lambda_{kl}{1\over r^{D-1\over 2}}\partial_r\,
r^{D-1\over 2}+{c_2[l]-\bar c_2+\delta_D\over 2r^2}-{1\over
r}\right)u_{kl}=0
\end{eqnarray}

Let $y_{kl}=r^{D-1\over 2}u_{kl}$, then the above equation becomes
\begin{eqnarray}
\left( {d^2\over dr^2} -2\lambda_{kl} {d\over dr}+\left[{2\over
r}-{c_2[l]-\bar c_2+\delta_D+{(D-1)(D-3)\over 4}\over
r^2}\right]\right)y_{kl}(r)=0\end{eqnarray}

Assume that $y_{kl}(r) \to r^s$ as $r\to 0^+$, then we must have the
following indicial equation:
\begin{eqnarray}\label{painleve}
\fbox{$s(s-1)=c_2[l]-\bar c_2+\delta_D+{(D-1)(D-3)\over 4}\,.$}
\end{eqnarray}

The further analysis is divided into two cases: 1) $D$ is odd, 2)
$D$ is even.

\subsection{The odd dimensional cases}
Let $D=2n+1$. Let $L^2({\cal S}^{2\mu}|_{\mr{S}^{2n}} )$ be the
$L^2$-sections of vector bundle ${\cal S}^{2\mu}$ restricted to the
unit sphere $\mr{S}^{2n}$. From the representation theory, we know
that
\begin{eqnarray}
L^2({\cal S}^{2\mu}|_{\mr{S}^{2n}})=\hat\bigoplus_{l\ge 0}{\ms R}_l
\end{eqnarray} where ${\ms R}_l$ is the irreducible representation space of
$\mr{Spin}(2n+1)$ with highest weight $(l+|\mu|, |\mu|, \cdots,
|\mu|)$. It is then clear that the Hilbert spaces of bound states is
\begin{eqnarray}\label{predecompodd}
{\ms H}=\hat\bigoplus_{l\ge 0}{\cal H}_l\end{eqnarray} with ${\cal
H}_l$ being a subspace of $L^2({\bb R}_+, r^{2n}\,dr)\otimes {\ms
R}_l$. Here $L^2({\bb R}_+, r^{2n}\,dr)$ is the $L^2$-space of
complex-valued functions on half-line $\bb R_+$ with measure
$r^{2n}dr$.

The value of the quadratic Casimir operator of $\mr{so}(2n)$ on
representation $\mb s^{2\mu}$ is
$$\bar c_2=n\mu^2+n(n-1)|\mu|.$$  The value of the quadratic
Casimir operator of $\mr{so}(2n+1)$ on ${\ms R}_l$ is
$$c_2[l]=l^2+2l(n+|\mu|-{1\over 2})+n\mu^2+n^2|\mu|.$$

Plugging the values for $\bar c_2$ and $c_2[l]$ into Eq.
(\ref{painleve}), we get
$$s(s-1)=(l+n+|\mu|)(l+n+|\mu|-1).$$
Therefore, $s=l+n+|\mu|$ or $s=1-l-n-|\mu|$. The solution
$s=1-l-n-|\mu|$ must be rejected; otherwise, the wave-functions
cannot be square integrable near $r=0$. Just as in solving the
hydrogen atom problem, with $s=l+n+|\mu|$, we continue the analysis
by setting
$$
y_{kl}=r^s\sum_{m=0}^\infty a_m r^m
$$
with $a_0=1$ and then get the recursive relation: for $m\ge 1$, one
has
\begin{eqnarray}
a_m\left( (m+s)(m+s-1)-s(s-1)\right)=\left( 1-
\lambda_{kl}(m+s-1)\right)a_{m-1}. \end{eqnarray} As it has been
demonstrated in Ref. \cite{Cohen77}, the power series solution must
be a polynomial solution; otherwise, the wave-function will not be
square integrable for $r$ near infinity. Therefore, we must have
$$
\lambda_{kl}={1\over k+s-1}={1\over k+l+n+|\mu|-1}
$$ and that leads to the energy spectrum
\begin{eqnarray}\label{pspecodd}
E_{kl}=-{1/2\over (k+l+n+|\mu|-1)^2}
\end{eqnarray}
where $k$ must be a positive integer, and an orthogonal
decomposition
$$
{\cal H}_l=\hat\bigoplus_{k=1}^\infty{\cal H}_{kl}
$$ with each of ${\cal H}_{kl}$ being isomorphic to ${\ms R}_l$ as
$\mr{Spin}(2n+1)$-modules. Therefore, in view of Eq.
(\ref{predecompodd}), we have an orthogonal decomposition of ${\ms
H}$ into energy eigen-states:
\begin{eqnarray}
{\ms H}=\hat\bigoplus_{I=0}^\infty\,{\ms H}_I
\end{eqnarray}
where
\begin{eqnarray*}
{\ms H}_I=\bigoplus_{k+l=I+1}\,{\cal H}_{kl}.
\end{eqnarray*}
Since linear action of $\mr{Spin}(2n+2)$ on ${\ms H}$ commutes with
the hamiltonian, this linear action must leave the energy
eigen-states ${\ms H}_I$ invariant. On the other hand, from
representation theory, as a $\mr{Spin}(2n+1)$-module, being
isomorphic to
$$\bigoplus_{l=0}^I{\ms R}_l\,,$$ ${\ms H}_I$ must be the irreducible
representation of $\hbox{Spin}(2n+2)$ with highest weight $(I+|\mu|,
|\mu|,  \cdots, |\mu|, \pm|\mu|)$. As a consistency check, one can
see that Eq. (\ref{hoperator}) yields
\begin{eqnarray}\label{gspecodd}
E_I=-{1/2\over (I+n+|\mu|)^2}
\end{eqnarray} on such representation, in complete agreement with Eq.
(\ref{pspecodd}) because $I=k+l-1$. One can show that, as a
$\mr{Spin}(2n+2)$-module, ${\ms H}_I$ has the highest weight
$(I+|\mu|, |\mu|,  \cdots, |\mu|,\mu)$.

\vskip 10pt In summary,  the energy spectrum is
\begin{eqnarray}
E_I=-{1/2\over (I+n+|\mu|)^2}
\end{eqnarray}
where $I=0$, $1$, $2$, \ldots; and $\ms H$ furnishes a
representation for $\mr{Spin}(2n+2)$ and has the following
decomposition into energy eigenstates:
\begin{eqnarray}
{\ms H}=\hat\bigoplus_{I=0}^\infty {\ms H}_I
\end{eqnarray} where $\ms H_I$ is the irreducible component of $\ms H$ with highest weight
$(I+~|\mu|, |\mu|, \cdots, |\mu|, \mu)$.

\subsection{The even dimensional cases}
Let $D=2n$. Let $L^2({\cal S}^{2\mu}|_{\mr{S}^{2n-1}} )$ be the
$L^2$-sections of vector bundle ${\cal S}^{2\mu}$ restricted to the
unit sphere $\mr{S}^{2n-1}$. From the representation theory, we know
that\footnote{For a fixed $l$, there are $(2\mu+1)$ many of $\ms
R_l$'s. When $\mu>1/2$, Eq. (\ref{errorsource}) below is no longer
valid for some ${\ms R}_l$ and the subsequent analysis fails. That
is the analytic reason for requiring $\mu=0$ or $1/2$.}
\begin{eqnarray}
L^2({\cal S}^{2\mu}|_{\mr{S}^{2n-1}} )=\left\{
\begin{array}{ll}
\hat\bigoplus_{l\ge 0} ({\ms R}_{l}^+\oplus{\ms R}_{l}^-) & \mbox{if $\mu=1/2$}\\
\\
\hat\bigoplus_{l\ge 0} {\ms R}_{l}^0 & \mbox{if $\mu=0$}
\end{array}\right.
\end{eqnarray} where ${\ms R}_{l}^\pm$ is the irreducible representation of
$\mr{Spin}(2n)$ with highest weight $(l+1/2, 1/2, \cdots, 1/2,\pm
1/2)$ and ${\ms R}_{l}^0$ is the irreducible representation of
$\mr{Spin}(2n)$ with highest weight $(l, 0, \cdots, 0)$. It is then
clear that the Hilbert spaces of bound states is
\begin{eqnarray}\label{wpredecompeven}
{\ms H}=\left\{
\begin{array}{ll}
\hat\bigoplus_{l\ge 0}({\cal H}_{l}^+\oplus {\cal H}_{l}^-) & \mbox{if $\mu=1/2$}\\
\\
\hat\bigoplus_{l\ge 0}{\cal H}_{l}^0 & \mbox{if
$\mu=0$}\end{array}\right.
\end{eqnarray}
with ${\cal H}_{l}^\sigma$ being a subspace of $L^2({\bb R}_+,
r^{2n-1}\,dr)\otimes {\ms R}_{l}^\sigma$. Here $L^2({\bb R}_+,
r^{2n-1}\,dr)$ is the $L^2$-space of complex-valued functions on
half-line $\bb R_+$ with measure $r^{2n-1}dr$.

The value of the quadratic Casimir operator of $\mr{so}(2n-1)$ on
representation $\mb s^{2\mu}$ is
$$\bar c_2=(n-1)\mu^2+(n-1)^2\mu.$$  The value of the quadratic
Casimir operator of $\mr{so}(2n)$ on ${\ms  R}_{l}^\sigma$ is
\begin{eqnarray}\label{errorsource}
c_2[l]=l^2+2l(n+\mu-1)+n\mu^2+(n^2-n)\mu.
\end{eqnarray}

Plugging the values for $\bar c_2$ and $c_2[l]$ into Eq.
(\ref{painleve}), we get
$$s(s-1)=(l+n+\mu-{1\over 2})(l+n+\mu-{3\over 2}).$$
Therefore, $s=l+n+\mu-{1\over 2}$ or $s={3\over 2}-l-n-\mu$. The
solution $s={3\over 2}-l-n-\mu$ must be rejected; otherwise, the
wave-functions cannot be square integrable near $r=0$. Just as in
solving the hydrogen atom problem, with $s=l+n+\mu-{1\over 2}$, we
continue the analysis by setting
$$
y_{kl}=r^s\sum_{m=0}^\infty a_m r^m
$$
with $a_0=1$ and then get the recursive relation: for $m\ge 1$, one
has
\begin{eqnarray}
a_m\left( (m+s)(m+s-1)-s(s-1)\right)=\left( 1-
\lambda_{kl}(m+s-1)\right)a_{m-1} .\end{eqnarray}As it has been
demonstrated in Ref. \cite{Cohen77}, the power series solution must
be a polynomial solution; otherwise, the wave-function will not be
square integrable for $r$ near infinity. Therefore, we must have
$$
\lambda_{kl}={1\over k+s-1}={1\over k+l+n+\mu-{3\over 2}}
$$ and that leads to the energy spectrum
\begin{eqnarray}\label{wpspeceven}
E_{kl}=-{1/2\over (k+l+n+\mu-{3\over 2})^2}
\end{eqnarray}
where $k$ must be a positive integer, and an orthogonal
decomposition
$$
{\cal H}_{l}^\sigma=\hat\bigoplus_{k=1}^\infty{\cal H}_{k{l}}^\sigma
$$ with each ${\cal H}_{k{l}}^\sigma$ being isomorphic to ${\ms R}_{l}^\sigma$ as
$\mr{Spin}(2n)$-modules. Therefore, in view of Eq.
(\ref{wpredecompeven}), we have an orthogonal decomposition of ${\ms
H}$ into energy eigen-states:
\begin{eqnarray}
{\ms H}=\hat\bigoplus_{I=0}^\infty\,{\ms H}_I
\end{eqnarray}
where
\begin{eqnarray*}
{\ms H}_I=\left\{ \begin{array}{ll} \bigoplus_{k+l=I+1}\,({\cal
H}_{kl}^+\oplus {\cal H}_{kl}^- ) & \mbox{if $\mu=1/2$}\\
\\
\bigoplus_{k+l=I+1}\,{\cal H}_{kl}^0  & \mbox{if $\mu=0$\;.}
\end{array}\right.
\end{eqnarray*}
Note that ${\ms H}_I$ is isomorphic to
$$\left\{ \begin{array}{ll}\bigoplus_{l=0}^I({\ms R}_{l}^+\oplus {\ms R}_{l}^-) & \mbox{if $\mu=1/2$}\\
\\\bigoplus_{l=0}^I{\ms R}_{l}^0 & \mbox{if $\mu=0$}\end{array}\right.$$
as a $\mr{Spin}(2n)$-module, from the representation theory, the
manifest $\mr{Spin}(2n)$ linear action on ${\ms H}_I$ can be
extended to a linear action of $\mr{Spin}(2n+1)$ such that ${\ms
H}_I$ is the irreducible representation of $\hbox{Spin}(2n+1)$ with
highest weight $(I+\mu, \mu, \cdots, \mu)$ and Eq. (\ref{hoperator})
is valid. As a consistency check, one can see that Eq.
(\ref{hoperator}) yields
\begin{eqnarray}\label{wgspecodd}
E_I=-{1/2\over (I+n+\mu-{1\over 2})^2}
\end{eqnarray} on such representation, in complete agreement with Eq.
(\ref{wpspeceven}) because $I=k+l-1$.

\vskip 10pt In summary,  the energy spectrum is
\begin{eqnarray}
E_I=-{1/2\over (I+n+\mu-{1\over 2})^2}
\end{eqnarray}
where $I=0$, $1$, $2$, \ldots; and $\ms H$ furnishes a
representation for $\mr{Spin}(2n+1)$ and has the following
decomposition into energy eigenstates:
\begin{eqnarray}
{\ms H}=\hat\bigoplus_{I=0}^\infty {\ms H}_I
\end{eqnarray} where $\ms H_I$ is the irreducible
component of $\ms H$ with highest weight $(I+~\mu, \mu, \cdots,
\mu)$.

\appendix
\section{Proof of the remaining part of Lemma \ref{lemma}}
The case $n=1$ is trivial. So we assume that $n\ge 2$. To prove
identity (\ref{specialodd}), we first note that we just need to
prove that identities
\begin{eqnarray}\label{sodd1}
\sum_k(\gamma_{1,k})^2 & = & {1\over n}c_2
\end{eqnarray}
and
\begin{eqnarray}\label{sodd2}
\sum_k\{\gamma_{1,k},\gamma_{2, k}\} & = &0
\end{eqnarray} hold in the representation $\mb s_+^{2\mu}$ for any non-negative half integer
$\mu$.

To continue, a digression on Lie algebra $\mr{so}(2n)$ is needed.
Recall that the root space of $\mr{so}(2n)$ is $\bb R^n$. Let
$e^i$ be the vector in $\bb R^n$ whose $i$-th entry is $1$ and all
other entries are zero. The positive roots are $e^i\pm e^j$ with
$1\le i<j\le n$. The simple roots are $\alpha^i=e^i-e^{i+1}$,
$i=1$ to $n-1$ and $\alpha^n=e^{n-1}+e^n$. For the Cartan basis,
we make the following choice: The commuting generators in the
Cartan subalgebra are taken to be
$$
H_j=-\gamma_{2j-1, 2j}\quad\quad \hbox{$j=1$ to $n$},
$$
and the $E$ generators are taken to be
\begin{eqnarray}
E_{\eta e^j+\eta' e^k}=-{1\over 2}\left(\gamma_{2j-1, 2k-1}+i\eta
\gamma_{2j, 2k-1}+i\eta'\gamma_{2j-1,
2k}-\eta\eta'\gamma_{2j,2k}\right)
\end{eqnarray} where $j<k$, and $\eta$, $\eta'$ $\in\{1,-1\}$.
Note that, the fact that
$$
[E_\alpha, E_\beta]=0\quad \hbox{\em if $\alpha+\beta$ is neither
a root nor zero}
$$ is frequently used in all subsequent calculations. All of these
are standard materials taken from a textbook such as Ref.
\cite{georgi82}.

Let \begin{eqnarray} \ms{O} & = &\sum_{n\ge
i\ge2}E_{-e^1-e^i}E_{-e^1+e^i}\\ \ms{O}_1 &=& H_1^2+{1\over
2}\sum_{n\ge i\ge 2}
\left(\{E_{-e^1-e^i},E_{e^1+e^i}\}+\{E_{-e^1+e^i},E_{e^1-e^i}\}\right)
\end{eqnarray}
By simple computations, we have
\begin{eqnarray}
\sum_k (\gamma_{1,k})^2 & = & \ms{O}_1+\ms{O}^\dag+\ms{O}\\
\sum_k \{\gamma_{1,k}, \gamma_{2,k}\} & = &{2\over
i}(\ms{O}^\dag-\ms{O})
\end{eqnarray}

It is then clear from the above calculations that identities
(\ref{sodd1}) and (\ref{sodd2}) are valid modulo the following
claim:
\begin{claim}
Let $|\Lambda\rangle$ be an element of the $\mr{so}(2n)$-module
$\mb s_+^{2\mu}$. Then
\begin{eqnarray}
\ms{O}|\Lambda\rangle & = &0,\\ \ms{O}^\dag|\Lambda\rangle
&=&0,\\
 \ms{O}_1|\Lambda\rangle &=& \mu(n+\mu-1)|\Lambda\rangle\\ & =&
 {1\over
n}c_2|\Lambda\rangle.
\end{eqnarray}
\end{claim}
\noindent {\bf {Proof of the claim}}. We first remark that, among
the four equalities in the claim, we just need to prove the first
and the third, that is because the 2nd is a consequence of the
first and the last is true because $c_2=n\mu(n+\mu-1)$. We also
remark that we may assume that $|\Lambda\rangle$ is a state
created from $|\mu,\cdots, \mu\rangle$ by applying a bunch of
lowing operators of the form $E_{-\alpha^j}$'s, that is because a
general state is always a linear combination of states of this
kind.

\vskip 10pt Next, we observe that
\begin{eqnarray}E_{-e^1+e^i}|\mu,\cdots, \mu\rangle=0;
\end{eqnarray}
consequently, $\ms{O}|\mu,\cdots, \mu\rangle=0$. This observation
can be shown by the following trick: Let $e=E_{e^1-e^i}$,
$f=E_{-e^1+e^i}$, $h=H_1-H_i$, then
\begin{eqnarray*}
[h,e] & = & 2e,\cr [h,f] &=& -2f,\cr [e,f]&=&h.
\end{eqnarray*} I.e., $\{e, f, h\}$ forms the
standard Cartan basis for $\mr{su}(2)$. It follows from the
following computation
\begin{eqnarray*}|| f|\mu,\cdots,
\mu\rangle||^2 =\langle \mu,\cdots, \mu |ef|\mu,\cdots, \mu\rangle =
\langle \mu,\cdots, \mu |h+fe|\mu,\cdots, \mu\rangle =0
\end{eqnarray*} that $f|\mu,\cdots, \mu\rangle=0$. Moreover,
when $|\Lambda\rangle =|\mu,\cdots, \mu\rangle$, the third
identity of the claim is just the consequence of a direct
computation. Therefore, the claim is true when $|\Lambda\rangle
=|\mu,\cdots, \mu\rangle$.

\vskip 10pt Finally, we need to reduce the general case to the
special case discussed in the previous paragraph. Combining the
computational fact that
\begin{eqnarray}
[E_{-e^1+e^j}, E_{-e^j+e^{j+1}}] =-iE_{-e^1+e^{j+1}},\;
[E_{-e^1-e^{j+1}},E_{-e^j+e^{j+1}}]= iE_{-e^1-e^j}
\end{eqnarray}
(where $1< j<n$) and the computational fact that
\begin{eqnarray}
[E_{-e^1+e^{n-1}}, E_{-e^{n-1}-e^n}] =-iE_{-e^1-e^n},\;
[E_{-e^1+e^n},E_{-e^{n-1}-e^n}]= iE_{-e^1-e^{n-1}},
\end{eqnarray}
one can show that $[\ms{O}, E_{-\alpha^j}]=0$ for any $1\le j\le
n$. Since $|\Lambda\rangle$ can be assumed to be a state created
from $|\mu,\cdots, \mu\rangle$ by applying a bunch of lowing
operators of the form $E_{-\alpha^j}$'s, in view of the fact that
$\ms{O}|\mu,\cdots, \mu\rangle =0$, we have $\ms{O}|\Lambda\rangle
=0$.

The reduction to the special case for the third identity is a bit
involved. The key is to introduce a series of operators:
\begin{eqnarray}\label{defO}
\ms{O}_2 & = & -2[\ms{O}_1, E_{-\alpha^1}]\\ \ms{O}_k & = &
i[\ms{O}_{k-1}, E_{-\alpha^{k-1}}]\quad\hbox{for $3\le k\le n$}\cr
\ms{O}^{n-1} & = & -i[\ms{O}_{n-1}, E_{-\alpha^n}]\cr \ms{O}^{k-1} &
= & -i[\ms{O}^k, E_{-\alpha^k}]\quad\hbox{for $1\le k\le
n-1$}\nonumber
\end{eqnarray}
and then observe that
\begin{eqnarray}\label{propO}
[\ms{O}_k, E_{-\alpha^j}] &=& 0\quad\hbox{ if $k\neq n-1$ and $j\neq
k$}\cr [\ms{O}_{n-1}, E_{-\alpha^j}] &=& 0\quad\hbox{ if $j\neq n-1,
n$}\cr [\ms{O}^k, E_{-\alpha^j}]  & = & 0 \quad\hbox{ if $j\neq
k$}\cr \ms{O}^0 &=& 4i\ms{O}\cr \ms{O}_k |\mu,\cdots,\mu\rangle &=&
0\quad\hbox{for $2\le k\le n$}\cr \ms{O}^k |\mu,\cdots,\mu\rangle
&=& 0\quad\hbox{for $0\le k\le n-1$}
\end{eqnarray}

With the help of the above equations starting from (\ref{defO}) and
ending at (\ref{propO}), an induction argument finishes the proof.
Here the induction is done on the number of lowing operators of the
form $E_{-\alpha^j}$'s which are used to create state
$|\Lambda\rangle$.

We end this appendix with the following conjecture: \emph{identity
(\ref{specialodd}) holds if and only if when the representation is a
Young power of $\mb s_+$ or $\mb s_-$.} For our purpose, we have
just proved the ``if" part in this appendix.

\end{document}